\documentstyle[11pt]{article}
\newcommand{\B}{{\cal B}}

\begin{document}
\begin{center}
{\LARGE{\bf Energy and Momentum Associated with  Solutions
Exhibiting Directional Type Singularities}}\\[2em]
\large{\bf{Ragab M. Gad}\footnote{Email Address: ragab2gad@hotmail.com}}\\
\normalsize {Mathematics Department, Faculty of Science,}\\
\normalsize  {Minia University, 61915 El-Minia,  EGYPT.}
\end{center}

\begin{abstract}

We obtain the energy and momentum densities of a general static
axially symmetric vacuum space-time, Weyl metric, with the help of
Landau-Lifshitz and Bergmann-Thomson  energy-momentum complexes.
We find that these two  definitions of energy-momentum complexes
do not provide the same energy density for the space-time under
consideration, while give the same momentum density. We show that,
in the case of Curzon metric (a particular case of the Weyl
metric), these two definitions give the same energy only when $R
\rightarrow \infty$. Furthermore, we compare these results with
those obtained using Einstein, Papapetrou  and M{\o}ller energy
momentum complexes.
\end{abstract}

%%% ----------------------------------------------------------------------
%\maketitle
%%% ----------------------------------------------------------------------
\setcounter{equation}{0}
\section{Introduction}
The notion of energy-momentum localization has been one of the
most interesting and thorny problems which remains unsolved since
the advent of general theory of relativity. Misner et al.
\cite{MTW} argued that the energy is localizable only for
spherical systems. Cooperstock and Sarracino \cite{CS}
contradicted their viewpoint and argued that if the energy is
localizable in spherical systems then it is also localizable for
all systems. Bondi \cite{Bo} expressed that a non-localizable form
of energy is inadmissible in relativity and its location can in
principle be found. In a series of papers, Cooperstock \cite{COO}
hypothesized that in a curved space-time energy and momentum are
confined  to the region of non-vanishing energy-momentum tensor
$T^{a}_{b}$  and consequently the gravitational waves are not
carriers of energy and momentum in vacuum space-times. This
hypothesis has neither been proved nor disproved. There are many
results support this hypothesis (see for example,
\cite{Xulu,Gad2}).  It would be interesting to investigate the
cylindrical gravitational waves in vacuum space-time. We use
Landau-Lifshitz and Bergmann-Thomson  energy-momentum complexes to
investigate whether or not these waves have energy and momentum
densities.
\par

The foremost endeavor to solve the problem of energy-momentum
localization was the energy-momentum complex introduced by
Einstein(E) \cite{E}. After this  many physicists, for instance,
Tolman (T) \cite{T}, Landau and Lifshitz (LL) \cite{LL},
Papapetrou (P) \cite{P}, Bergmann (B) \cite{B} and Weinberg (W)
\cite{W} (abbreviated to (ETLLPBW), in the sequel) have given
different definitions for the energy-momentum complexes. The major
difficlty with these attempts was that energy-momentum complexes
had to be computed in quasi-Cartesian coordinates. M{\o}ller (M)
\cite{E1} introduced a consistent expression which enables one to
evaluate energy and momentum in any coordinate system. Although of
these drawbacks, some interesting results obtained recently lead
to the conclusion that these energy-momentum complexes give the
same energy distribution for a given space-time
\cite{V1}-\cite{V8}. Aguirregabiria, Chamorro and Virbhadra
\cite{V9} showed that the five different energy-momentum complexes
(ELLPBW)  give the same result for the energy distribution for any
Kerr-Schild metric. Recently, Virbhadra \cite{V99} investigated
whether or not these definitions (ELLPBW) lead to the same result
for the most general non-static spherically symmetric metric and
found that they disagree. He noted that the energy-momentum
complexes (LLPW) give the same result as in the Einstein
prescription if the calculations are performed in Kerr-Schild
Cartesian coordinates. However, the complexes (ELLPW) disagree  if
computations are done in ``Schwarzschild Cartesian coordinates
\footnote{ Schwarzschild metric in ``Schwarzschild Cartesian
coordinates'' is obtained by transforming this metric (in usual
Schwarzschild coordinates $\{t, r, \theta, \phi\}$)  to
$\{t,x,y,z\}$ using $ x = r \sin\theta \cos\phi, y = r \sin\theta
\sin\phi, z = r \cos\theta $.}".
\par
Some interesting results \cite{Xulu1}-\cite{Gad3} led to the
conclusion that in a given space-time, such as: the
Reissner-Nordst\"{o}rm, the de Sitter-Schwarzschild, the charged
regular metric, the stringy charged black hole and the
G\"{o}del-type space-time, the energy distribution according to
the energy-momentum complex of Einstein is different from that of
M{\o}ller. But in some specific case \cite{E1,Xulu1,V97,V99} (the
Schwarzschild, the Janis-Newman-Winicour metric) have the same
result.

\par
The scope of this paper is to evaluate  the energy  and momentum
densities for the solutions exhibiting directional singularities
using Landau-Lifshitz and Bergmann-Thomson energy-momentum
complexes. In general relativity the term "directional
singularity" is applied if the limit of an invariant scalar
(Kretschmann scalar ${\bf{{\mathcal{K}} = R_{abcd}R^{abcd}}}$,
$\bf{{R_{abcd}}}$ are the components of the Riemann tensor)
depends upon the direction by which the singularity is approached.
One of the best known examples of such directional behavior is the
Curzon singularity occurring at $R = 0$ in the Weyl metric
\cite{11}. Gautreau and Anderson \cite{GA} showed that for the
field of a Curzon \cite{Curzon} particle, the Kretschmann scalar
${\bf{{\mathcal{K}}}}$ tends to the value zero along the z-axis
but becomes infinite for other straight line trajectory to the
origin. A more detailed analysis ecoompassing a wider class of
curves was carried out by Cooperstock and Junevicus \cite{CJ}.

Through this paper we use $G = 1$ and $c = 1$ units and follow the
convention that Latin indices take value from 0 to 3 and Greek
indices take value from 1 to 3.
\par
The general static axially symmetric vacuum solution of
Einstein's field equations is given by the
Weyl metric \cite{11}
\begin{equation} \label{1}
ds^2 = e^{2\lambda} dt^2 - e^{2(\nu -\lambda )}(dr^2 + dz^2) - r^2 e^{-2\lambda} d\phi^2
\end{equation}
where
$$
\lambda_{rr} + \lambda_{zz} + r^{-1}\lambda_{r} = 0
$$
and
$$
\nu_{r} = r(\lambda^{2}_{r} - \lambda^{2}_{z}), \qquad \nu_{z} = 2r\lambda_{r}\lambda_{z}.
$$

\par
It is well known that if the calculations are performed in
quasi-Cartesian coordinates,
all the energy-momentum complexes give meaningful results.\\
According to the following transformations
$$
r = \sqrt{x^2 + y^2} , \qquad \phi = \arctan(\frac{y}{x}),
$$
the line element (\ref{1}) written in terms of  quasi-Cartesian
coordinates reads:
$$
ds^2 = e^{2\lambda}dt^2 - \frac{1}{r^2}\big( x^2e^{2(\nu -
\lambda)} + y^2e^{-2\lambda} \big) dx^2 -
\frac{2xy}{r^2}\Big(e^{2(\nu - \lambda)} - e^{-2\lambda}\Big)dxdy-
$$
\begin{equation} \label{2}
\frac{1}{r^2} \big( y^2e^{2(\nu - \lambda)} +
x^2e^{-2\lambda}\big)dy^2  -  e^{2(\nu - \lambda)}dz^2,
\end{equation}
where
$$
x^2\lambda_{xx} + y^2\lambda_{yy} + 2xy\lambda_{xy} +
r^2\lambda_{zz} + x\lambda_{x} + y\lambda_{y} = 0,
$$
$$
x\nu_{x} + y\nu_{y} -(x\lambda_{x} + y\lambda_{y})^2
+r^2\lambda_{z} = 0
$$
and
$$
\nu_{z} = 2\lambda_{z}(x\lambda_{x} + y\lambda_{y}).
$$
\par
 For the above metric the determinant of the metric tensor
and the contravariant components of the tensor are given,
respectively, as follows
\begin{equation}\label{3}
%\begin{split}
\begin{array}{ccc}
det(g) & = - e^{4(\nu - \lambda)},\\
g^{00} & = e^{-2\lambda}, \\
g^{11} & = - \frac{e^{2\lambda }}{r^2} (y^2 + x^2 e^{-2\nu}),\\
g^{12} & =  \frac{xye^{2\lambda}}{r^2}(1-e^{-2\nu}),\\
g^{22} & = -\frac{e^{2\lambda}}{r^2}(x^2+y^2e^{-2\nu}),\\
g^{33} & = - e^{2(\lambda - \nu)}.
%\end{split}
\end{array}
\end{equation}
\setcounter{equation}{0}
\section{\bf{Energy-momentum Complex in Landau-Lifshitz's Prescription}}
The energy-momentum complex of Landau and Lifshitz \cite{LL} is
\begin{equation}\label{ll1}
L^{ij}= \frac{1}{16\pi}S^{ikjl}_{\quad ,kl},
\end{equation}
where $S^{ikjl}$ with symmetries of the Riemann tenor and is
defined by
\begin{equation}\label{LL2}
S^{ikjl}=-g(g^{ij}g^{kl}-g^{il}g^{kj}).
\end{equation}
The quantity $L^{00}$ represents the energy density of the whole
physical system including gravitation and $L^{0\alpha}$ represents
the components of the total momentum (energy current) density.
\par
In order to evaluate the energy and momentum densities in
Landau-Lifshitz's prescription associated with the Weyl metric
(\ref{1}), we evaluate the non-zero components of $S^{ikjl}$

\begin{equation}\label{3.5}
%\begin{split}
\begin{array}{ccc}
S^{0101} & = -\frac{e^{4(\nu - \lambda)}}{r^2}\big(y^2 + x^2 e^{-2\nu}\big),\\
S^{0102} & = \frac{xye^{4(\nu - \lambda)}}{r^2}\big(1-
e^{-2\nu}\big),\\
S^{0202} & = -\frac{e^{4(\nu - \lambda)}}{r^2}\big(x^2 + y^2
e^{-2\nu}\big),\\
S^{0303} & = -e^{2\nu - 4\lambda}.
%\end{split}
\end{array}
\end{equation}
Using these components  in equation (\ref{ll1}),  we get the
energy and momentum densities as following
$$
\begin{array}{ccc}
 L^{00}& =
-\frac{1}{8\pi r^2}e^{2\nu -4\lambda}\Big[x^2\nu_{xx} +
y^2\nu_{yy} + 2xy\nu_{xy} - 8y^2\nu_{y}\lambda_{y}
-8x^2\nu_{x}\lambda_{x} + r^2\nu_{zz}+
\\& 8(x\lambda_{x}+y\lambda_{y})^2 + 2(x\nu_{x}+y\nu_{y})^2 +
2r^2(\nu_{z}-2\lambda_{z})^2 -8xy\nu_{y}\lambda_{x} -
8xy\nu_{x}\lambda_{y} +\\
&2(x\nu_{x}+y\nu_{y}) - 2(x\lambda_{x}+y\lambda_{y})+
e^{2\nu}\big( 2y^2\nu_{xx} + 2x^2\nu_{yy} - 4xy\nu_{xy} -
2y^2\lambda_{xx} - \\&2x^2\lambda_{yy} + 4xy\lambda_{xy}
-16xy\nu_{x}\nu_{y} + 16xy\lambda_{y}\nu_{x} -
16xy\lambda_{y}\lambda_{x} +  16xy\nu_{y}\lambda_{x} +\\&
8y^2(\nu_{x}-\lambda_{x})^2 + 8x^2(\nu_{y}-\lambda_{y})^2 +
4(x\lambda_{x}+y\lambda_{y}) - 4(x\nu_{x}+y\nu_{y})\big)\Big],
\end{array}
$$
in the cylindrical polar coordinates the energy density takes the
form
$$
L^{00}=-\frac{1}{8\pi r^2}e^{2\nu
-4\lambda}\Big[r^2\nu_{rr}+2r^2(\nu_{r}-2\lambda_{r})^2
+r^2\nu_{zz} + 2r^2(\nu_{z}-2\lambda_{z})^2 - 2r(\nu_{r} -
\lambda_{r})(e^{2\nu}-1)\Big],
$$
$$
L^{\alpha 0} = 0.
$$
The momentum components are vanishing everywhere.
\par
We now restrict our selves to the particular solutions of  Curzon
metric \cite{Curzon} obtained by setting
$$
\lambda = - \frac{m}{R} \qquad  and \quad \nu = -\frac{m^2
r^2}{2R^4}, \qquad R = \sqrt{r^2 + z^2}
$$
in equation (\ref{1}). \\
For this solution  the energy and momentum densities  become
\begin{equation}\label{3.9}
\begin{array}{ccc}
L^{00}& = \frac{1}{8\pi}e^{4(\nu -
\lambda)}\Big[-\frac{4m^2r^2}{R^6} - \frac{2m^2}{R^4}
-\frac{4m}{R^3}+ \\&e^{-2\nu}\big( -\frac{5m^2}{R^4} -
\frac{4m^2r^2}{R^6} +\frac{4m}{R^3} - \frac{2m^4r^2}{R^8} +
\frac{8m^3r^2}{R^7}\big)\Big],
\end{array}
\end{equation}
\begin{equation}\label{3.10}
L^{\alpha 0} = 0.
\end{equation}
The momentum components are vanishing everywhere.

\setcounter{equation}{0}
\section{\bf{The Energy-Momentum Complex of Bergmann-Thomson}}
The Bergmann-Thomson  energy-momentum complex \cite{B} is given by
\begin{equation}\label{6.1}
B^{ik} = \frac{1}{16\pi}\big[g^{il}\B^{km}_{l}\big]_{,m},
\end{equation}
where
$$
\B^{km}_{l} =
\frac{g_{ln}}{\sqrt{-g}}\Big[-g\Big(g^{kn}g^{mp}-g^{mn}g^{kp}\Big)\Big]_{,p}.
$$
$B^{00}$ and $B^{0\alpha}$ are the energy and momentum density
components. In order to calculate $B^{00}$ and $B^{0\alpha}$ for
Weyl metric, using Bergmann-Thomson  energy-momentum complex, we require
the following non-vanishing components of $H^{km}_{l}$
\begin{equation}\label{6.5}
\begin{array}{ccc}
\B^{01}_{0} & = \frac{1}{r^2}\big[ 2x^2(2\lambda_{x}-\nu_{x}) +
2xy(2\lambda_{y}-\nu_{y})+4xye^{2\nu}(\nu_{y}-\lambda_{y})+\\
&x(e^{2\nu}-1) + 4y^2e^{2\nu}(\lambda_{x}-\nu_{x})\big]\\
\B^{02}_{0} & = \frac{1}{r^2}\big[2y^2(2\lambda_{y}-\nu_{y}) +
2xy(2\lambda_{x}-\nu_{x}) + 4xye^{2\nu}(\nu_{x}-\lambda_{x})\\
& y(e^{2\nu}-1) + 4x^2e^{2\nu}(\lambda_{y}-\nu_{y})\big]\\
 \B^{03}_{0} & = 2(2\lambda_{z}-\nu_{z}).
\end{array}
\end{equation}

 Using
the components (\ref{6.5}) and (\ref{3}) in (\ref{6.1}), we get
the energy and momentum densities for the Weyl metric,
respectively, as follows
$$
B^{00} = \frac{e^{-2\lambda}}{8\pi r^2}\Big[-x^2\nu_{xx} -
y^2\nu_{yy}-2xy\nu_{xy}-4(x\lambda_{x}+y\lambda_{y})^2+(y\lambda_{y}+x\lambda_{x})
$$
$$
+(y\nu_{y}
+x\nu_{x})+2(x\nu_{x}+y\nu_{y})(x\lambda_{x}+y\lambda_{y})-2r^2\nu_{z}\lambda_{z}
+2r^2\nu_{z}\lambda_{z}-4r^2\lambda_{z}^{2}
$$
$$
+ e^{2\nu}\Big(2x^2\lambda_{yy}+2y^2\lambda_{yy}-4xy\lambda_{xy}
-2y^2\nu_{xx}-2x^2\nu_{yy}+4xy\nu_{xy}
$$
$$
+8y\lambda_{x}(y\nu_{x}-x\nu_{y})+8x\lambda_{y}(y\nu_{x}-x\nu_{y})-4(x\nu_{y}-y\nu_{x})^2
-4(y\lambda_{x}-x\lambda_{y})^2
$$
\begin{equation}\label{6.6}
+3(y\nu_{y}+x\nu_{x})-3(y\lambda_{y}+x\lambda_{x})\Big)\Big].
\end{equation}
\begin{equation}\label{6.7}
B^{0\alpha} = 0.
\end{equation}
The momentum components are vanishing everywhere.
\par
Using cylindrical polar coordinates the energy density takes the
form
$$
B^{00} = \frac{e^{-2\lambda}}{8\pi r^2}\Big[
2r^2\nu_{r}\lambda_{r}-3r^2(\lambda_{r}^{2}+\lambda_{z}^{2})+2r^2\nu_{z}\lambda_{z}
-r(e^{2\nu}-1)(\lambda_{r}-\nu_{r})\Big].
$$
 For the Curzon solution, using equations (\ref{6.5}) and (\ref{6.6}),
 the energy and momentum densities become
$$
B^{00} = \frac{m}{8\pi R^3}e^{-2\lambda}\Big[-\frac{2m}{R} +
\frac{2m^2r^2}{R^4} -(e^{2\nu} - 1) - \frac{2mr^2}{R^3}-
$$
\begin{equation}\label{6.8}
e^{2\nu}\big(\frac{m}{R}-\frac{2mr^2}{R^3}\big)\Big]
\end{equation}
\begin{equation}\label{6.9}
B^{0\alpha} = 0.
\end{equation}
The momentum components are vanishing everywhere.

\par
 In the following table we summarize our results
obtained  (see, \cite{Gad4}) of the energy and momentum densities
for Curzon metric, using  Einstein, Papapetrou and M{\o}ller.
\newpage

\begin{table}
  \centering
  \begin{tabular}{|c|c|c|}
  %\begin{tabular}{|t|l|}
\hline
    {\bf{Prescription}}& {\bf{Energy density}} & {\bf{Momentum density}} \\
     \hline
Einstein & $\theta^{0}_{0} =
\frac{1}{16\pi}\Big[-\frac{4m^2r^2}{R^6} +
\frac{4m^2}{R^4} + $&\\
     &
$2e^{2\nu}\big( -\frac{m^2}{R^4} +
\frac{2m^2r^2}{R^6}\big)\Big]$ & $\theta_{\alpha}^{0} = 0$.\\
 \hline
    Papapetrou & $\Omega^{00} = \frac{1}{16\pi}\Big[-e^{2\nu - 4\lambda}\big(
\frac{4m^4r^2}{R^8} + \frac{12m^2}{R^4} - \frac{16m^3r^2}{R^7} +$&\\
& $\frac{4m^2}{R^6}\Big) +
2e^{2\nu}\Big(\frac{2m^2r^2}{R^6}-\frac{m^2}{R^4} \Big)\Big]$ &$\Omega^{\alpha 0} = 0$. \\
    \hline
     M{\o}ller & $\Im^{0}_{0} = 0$ & $\Im_{\alpha}^{0} = 0$.\\
\hline
  \end{tabular}
  \vspace{2mm}
  \caption{\sf{The energy and momentum densities, using (EPM),
   for the Curzon metric}}
\end{table}

\section*{\bf{Discussion}}
Using different definitions of energy-momentum complex, several
authors  studied the energy distribution for a given space-time.
Most of them restricted their intention to the static and
non-static spherically symmetric space-times. Rosen \cite{R}
calculated the energy and momentum densities of a non-static
cylindrical space-time using the energy-momentum pseudo tensors of
Einstein and Landau-Lifshitz. He found, if the calculations  are
preformed in cylindrical polar coordinates, that the energy and
momentum density components vanish. When the calculations are
carried out in Cartesian coordinates, Rosen and Virbhadra
\cite{RV} evaluated these quantities using Einstein's prescription
and found that these quantities turn out to be non-vanishing and
reasonable. Virbhadra \cite{V5} used Tolman, Landau-Lifshitz and
papapetrou's prescriptions and found that they give the same
energy and momentum densities for the aforementioned space-time
and agree with the results obtained by using Einstein's
prescription.
\par
In our previous two papers \cite{Gad4} we have calculated the
energy and momentum densities associated with a general static
axially symmetric vacuum space-time, using Einstein, Papapetrou
and M{\o}ller's energy-momentum complexes. We found that these
definitions do not provide the same energy density, while give the
same momentum density.
\par
 In this paper, we
calculated the energy and momentum density components for this
space-time using Landau-Lifshitz and Bergmann-Thomson
energy-momentum complexes. Further, using these results we
obtained the energy and momentum densities for the Curzon
metric.\\
We found that for both, Weyl and Curzon metrics, the
Landau-Lifshitz and Bergmann-Thomson give exactly the same
momentum density but do not provide the same energy density,
except only at $R \rightarrow \infty$, in the case of Curzon
metric, where the energy density tends to zero.
\par
Furthermore, we have made a comparison of our results with those
calculated \cite{Gad4} using (EPM) prescriptions. We obtained that
the five prescriptions (ELLPBM) give the same result regarding the
momentum density associated with Weyl as well as Curzon metrics.
Concerning the energy density associated with both two metrics
under consideration, we found that these prescriptions (ELLPBM) do
not give the same result except when $R \rightarrow \infty$, in
the case of Curzon metric, where the energy in  all
prescriptions (ELLPBM) tends to zero.\\
Finally, in the case of Curzon metric we see that the energy in
the  prescriptions (ELLPB) diverges at the singularity ($R = 0$),
but it will never diverge in M{\o}ller's prescription.
\vspace*{2mm}

%\begin{thebibliography}{widest-label}

\end{document}